\documentclass[12pt]{article}
\usepackage{graphicx}
\usepackage{times}
\usepackage[sort&compress]{natbib}
\usepackage{fancyhdr}
\usepackage{epsfig}
\usepackage{multirow}

\date{13 March 2011}
\begin{document}

\bibliographystyle{nature}

\renewcommand{\multirowsetup}{\centering}

\title{Phase transitions in contagion processes mediated by recurrent mobility patterns}

\author{Duygu Balcan$^{1,2}$ and  Alessandro Vespignani$^{1,2,3}$\footnote{To whom correspondence should be addressed; email: alexv@indiana.edu}}
\maketitle

\begin{center}
\small{
$^1$Center for Complex Networks and Systems Research (CNetS), School of Informatics and Computing, Indiana University, Bloomington, IN 47408, USA\\
$^2$Pervasive Technology Institute, Indiana University, Bloomington, IN 47406, USA\\
$^3$Institute for Scientific Interchange (ISI), Torino 10133, Italy \\}
\end{center}

\noindent {\bf
Human mobility and activity patterns mediate contagion on many levels, including the spatial spread of infectious diseases, diffusion of rumors, and emergence of consensus. These patterns however are often dominated by specific locations and recurrent flows and poorly modeled by the random diffusive dynamics generally used to study them. Here we develop a theoretical framework to analyze contagion within a network of locations where individuals recall their geographic origins. We find a phase transition between a regime in which the contagion affects a large fraction of the system and one in which only a small fraction is affected. This transition cannot be uncovered by continuous deterministic models due to the stochastic features of the contagion process and defines an invasion threshold that depends on mobility parameters, providing guidance for controlling contagion spread by constraining mobility processes.  We recover the threshold behavior by analyzing diffusion processes mediated by real human commuting data.}

In recent years, reaction-diffusion processes have been used as a successful modeling framework to approach a wide array of systems that along with the usual chemical and physical phenomena~\cite{Marro1999, vanKampen1981} includes epidemic spread~\cite{May1984, Bolker1993, Bolker1995, Sattenspiel1995, Lloyd1996, Keeling2002, Watts2005}, human mobility~\cite{Bolker1995, Sattenspiel1995, Lloyd1996, Keeling2002}, information, and social contagion processes~\cite{Rapoport1953, Goffman1964, Goffman1966, Dietz1967, Tabah1999, Daley2000}. This has stimulated the broadening of reaction-diffusion models in order to deal with complex network substrates and complex mobility schemes~\cite{Rvachev1985, Grais2003, Hufnagel2004, Colizza2006a, Balcan2010}. This success has allowed for the theoretical characterization of new and interesting dynamical behaviors and provide a rationale for the understanding of the emerging critical points that underpin some of the most interesting characteristics of techno-social systems. Those studies however are all focused on mobility processes modeled through simple memoryless diffusive processes. The recent accumulation of large amounts of data on human mobility~\cite{Chowell2003, Barrat2004, Guimera2005, Brockmann2006, Patuelli2007, Gonzalez2008} from the scale of single individuals to the scale of entire populations presents us with new challenges related to the high level of predictability and recurrence~\cite{Wang2009, Song2010a, Song2010b} found in mobility and diffusion patterns from real data. For instance, commuting mobility denoted by recurrent bidirectional flows among locations dominates by an order of magnitude the human mobility network at the scale of census areas defined by major urban areas~\cite{Balcan2009-PNAS}. The effect of highly-predictable or recurrent features of particles/agents mobility in the large-scale behavior of contagion processes however cannot be studied by a simple adaptation of previous theoretical frameworks~\cite{Colizza2007NatPhys, Colizza2007PRL, Colizza2008JTB, barthelemy2010,Ni2009, Ben-Zion2010} and call for specific methodologies and approximations capable of coping with non-markovian diffusive processes in complex networks. 

\noindent{\bf Modeling commuting networks}.

\noindent In order to start investigating the effect of regular mobility patterns in reaction-diffusion systems we have considered the prototypical example of the spread of biological agents and information processes in populations characterized by bidirectional commuting patterns. In this case we consider a system made of $V$ distinct subpopulations.  The $V$ subpopulations form a network in which each subpopulation $i$ has a population made of $N_i$ individuals and is connected to a set of other subpopulations $\upsilon(i)$. The edge connecting two subpopulations $i$ and $j$ indicates the presence of a flux of commuters. We assume that individuals in the subpopulation $i$ will visit anyone of the connected subpopulations with a per capita diffusion rate $\sigma_i$. As we aim at modeling commuting processes in which individuals have a memory of their location of origin, displaced individuals return to their original subpopulation with rate $\tau^{-1}$. 

Real data from commuting networks add an extra layer of complexity to the problem. In Fig. 1 we display the cumulative distributions of the number of commuting 
connections per administrative unit and the daily flux of commuters on each connection in the United States and France. The networks exhibit important variability in the number of connections per geographic area. Analogously, the daily number of commuters on each connection is highly heterogeneous, distributed in a wide range of four to six orders of magnitude. These properties, often mathematically encoded in a heavy-tailed probability distribution, have been shown to have important consequences for dynamical processes, altering the threshold behavior and the associated dynamical phase transition~\cite{Pastor-Satorras2001, lloyd2001, Cohen2003, Barrat2008, Colizza2007NatPhys, Colizza2007PRL, Colizza2008JTB}. In order to take into account the effect of the network topology we use a particle-network framework in which we consider a random subpopulation network with given degree distribution $P(k)$ and denote the number of subpopulations with $k$ connections by $V_k$. Furthermore, we assume statistical equivalence for subpopulations of similar degree. This is a mean-field approximation that considers all subpopulations with a given degree $k$ as statistically equivalent, thus allowing the introduction of degree-block variables that depend only upon the subpopulation degree~\cite{Colizza2008JTB}. While this is an obvious approximation of the system description, it has been successfully applied to many dynamical processes on complex networks and it is rooted in the empirical evidence gathered in previous works~\cite{Colizza2008JTB, Chowell2003, Barrat2004, Guimera2005}. For the sake of analysis we will assume that the average population in each node of degree $k$ follows the functional form $N_k={\overline N} k / \langle k \rangle$ where ${\overline N}=\sum_k N_k P(k)$  is the average number of individuals per node in the subpopulation network. This expression represents the stationary population distribution in the case of a simple random diffusive process in which the diffusion rate of individuals along each link leaving a node of degree $k$ has the form $1/ k$ \cite{Colizza2007PRL, Colizza2008JTB}. Moreover, the empirical data from various sources suggest similar population scaling arise as a function of their connectivity to other populations \cite{Barrat2004,Guimera2005,Colizza2006a}. 

In order to approach the spreading process in the subpopulation network analytically, we define mixing subpopulations~\cite{Sattenspiel1995,Keeling2002} that identify the number of individuals $N_{kk'}(t)$ of the subpopulation $k$ present in subpopulation $k'$ at time $t$ (see Fig. 2). We consider that the diffusion rate $\sigma_{kk'}$ is a function of the degree $k$ and $k'$ of the origin and destination subpopulations, respectively, with  $\sigma_k=\sum_{k' \in \upsilon(k)} \sigma_{kk'}$  and $\tau_k$ depending only on the degree of the origin subpopulation. In particular, if $\sigma_k \ll \tau_k^{-1}$ and we study the system on a time scale larger than the time scale of the commuting process $\tau_k$ one can consider a quasi-stationary approximation in which the mixed subpopulations assume their stationary values: 
\begin{equation}
N_{kk}=\ \frac{\overline{N} k}{\langle k\rangle (1+\sigma_k \tau_k)} 
\;\;\;,\label{eq:N_kk}
\end{equation}
\begin{equation}
N_{kk'}= \frac{\overline{N} k \sigma_{kk'} \tau_k}{\langle k\rangle (1+\sigma_k\tau_k) } 
\;\;\;.\label{eq:N_kk'}
\end{equation}
These expressions (see the Methods section) allow us to consider the subpopulation $k$ as if it had an effective number of individuals $N_{kk'}\ll N_{kk}$ in contact with the individuals of the neighboring subpopulation $k'$ in a quasi-stationary state reached whenever the time scale of the dynamical process we are studying is larger than $\tau_k$. For the sake of the analytical treatment in the following we will consider in the commuting rates only the dependence on the degree classes. More complicated functional forms including explicitly the spatial distance may be considered and we will analyze this case by performing data-driven simulations.

\noindent{\bf Contagion processes and the invasion threshold}.

\noindent In analyzing contagion processes in this system we consider the usual susceptible-infected-recovered (SIR) contagion model~\cite{Keeling2008}. Within each subpopulation the total number of individuals is partitioned into the compartments $S(t)$, $I(t)$ and $R(t)$, denoting the number of susceptible, infected, and removed individuals at time $t$, respectively. The basic SIR rules thus define a reaction scheme of the type $S+I\to 2I$ with reaction rate $\beta$ and $I\to R$ with reaction rate $\mu$, which represent the contagion and recovery processes, respectively. The SIR epidemic model conserves the number of individuals and is characterized by the reproductive number $R_0=\beta/\mu$ that determines the average number of infectious individuals generated by one infected individual in a fully-susceptible population. The epidemic is able to generate a number of infected individuals larger than those who recover only if $R_0>1$, yielding the classic result for the epidemic threshold~\cite{Keeling2008}; if the spreading rate is not large enough to allow a reproductive number larger than one (i.e., $\beta>\mu$), the epidemic outbreak will affect only a negligible portion of the population and will die out in a finite amount of time. 

While this result is valid at the level of each subpopulation, each subpopulation may or may not transmit the infection or contagion process to another subpopulation it is in contact with, depending on the level of mixing among the subpopulations. In other words, the mobility parameters $\sigma_k$ and $\tau_k$ influence the probability that individuals carrying infection or information will export the contagion process to nearby subpopulations. If the diffusion rate approaches zero the probability of the contagion entering neighboring subpopulations goes to zero as there are no occasions for the carriers of the process to visit them. On the other hand if the return rate is very high, then the visiting time of individuals in neighboring populations is so short that they do not have time to spread the contagion in the visited subpopulations. This implies the presence of a transition~\cite{Colizza2007PRL, Colizza2008JTB, Ball1997, Cross2005, Cross2007} between a regime in which the contagion process may invade a macroscopic fraction of the network and a regime in which it is limited to a few subpopulations (see Fig. 2 for a pictorial illustration).  In this perspective we can consider the subpopulation network in a coarse-grained view and provide a characterization of the invasion dynamics at the level of subpopulations, translating epidemiological and demographic parameters into Levins-type parameters of extinction and invasion rates. Let us define $D^0_k$ as the number of subpopulations of degree $k$ affected by the contagion at generation $0$, i.e., those which are experiencing the outbreak at the beginning of the process. Each subpopulation invaded by the contagion process will seed~--~during the course of the outbreak~--~the  contagion process in neighboring subpopulations, defining the set $D^1_k$ of invaded subpopulations at  generation  1, and so on. This corresponds to a basic branching process~\cite{Colizza2007PRL, Colizza2008JTB, Ball1997, Harris1989, Vazquez2006} where the $n$th generation of infected subpopulations of degree $k$ is denoted by $D^n_k$.  In order to describe the early stage of the subpopulation invasion dynamics we assume that the number of subpopulations affected by a contagion outbreak (with $R_0>1$) is small and we can therefore study the evolution of the number of subpopulations affected by the contagion process by using a tree-like approximation relating $D^n_k$ with $D^{n-1}_k$. As it is shown in the Methods section, in the case of  $R_0 \simeq 1$, it is possible to derive the following recursive equation
\begin{equation}
D_{k}^{n}=(R_{0}-1)\frac{kP(k)}{\langle k \rangle}\sum_{k'}D_{k'}^{n-1}(k'-1) \lambda_{k'k}\;\;\;.\label{branch}
\end{equation}
This relation carries explicit dependence on the network topology through the degree distribution $P(k)$ and the factor $\lambda_{k'k}$ that is the number of contagious seeds that are introduced into a fully-susceptible population of degree $k$ from a neighboring population of degree $k'$.  If the time scale of the disease is considerably larger than the commuting time scale, that is in our case  $\mu^{-1}\gg \tau$, we can consider the infectious individuals in the mixing subpopulation to assume their stationary values according to Eq.~(\ref{eq:N_kk'}). The quantity $\lambda_{k'k}$ can therefore be expressed as the total number of infected individuals in the mixing subpopulation by $\lambda_{k'k}=\left(N_{k'k}+N_{kk'}\right)\alpha$,  where $\alpha$ is the fraction of individuals that are affected by the contagion by the end of the SIR epidemic. The first term in the right-handside of the above expression accounts for the total visits of infectious people from source subpopulation $k'$ to target subpopulation $k$. While the second term counts for the visits of individuals from the target subpopulation to the source subpopulation, during which they acquire infection and carry the contagion back to their origin. If we use the steady state expression in equation (\ref{eq:N_kk'}), and we consider that  $\alpha$ for the SIR dynamics can be explicitly written for $R_0\simeq1$ it is possible to write an explicit form of the iterative equation (\ref{branch}), whose dynamical behavior is determined by the branching ratio
\begin{equation}
R_{\ast} = \frac{2 \overline{N} (R_{0}-1)^2 \rho }{R_{0}^{2} (1+  \langle k \rangle / \langle k^2 \rangle + \rho )} F(\langle k\rangle, \langle k^{2}\rangle, \langle k^{3}\rangle,\langle k^{4}\rangle)  
\;\;\;, \label{eq:InvasionThr1}
\end{equation}
where $\rho \equiv \sigma \tau$ is the ratio of commuting to return rate and for the sake of simplicity we have considered that the per capita commuting rate $\sigma$ and return rate $\tau^{-1}$ are the same for all subpopulations. In the above expression $F$ is a function only of the moments of the degree distribution of the subpopulation network. $R_\ast$ is therefore equivalent to a basic reproductive number at the subpopulation level, defining the average number of supopulations to which each infected subpopulation will spread the contagion process. $R_\ast$ thus defines the  invasion threshold as any contagion process will spread globally in the network system only if $R_{\ast}>1$. The subpopulation branching process is inherently considering the stochastic effects of the epidemic dynamics in the probability of contagion from one subpopulation to the other. It is  interesting to note that the invasion threshold cannot indeed be derived in continuous deterministic models where stochastic effects are neglected. 

\noindent{\bf Phase diagram and the network structure}.

\noindent  For fixed disease and network parameters, the condition $R_\ast=1$ of Eq.~(\ref{eq:InvasionThr1}) defines critical value for $\rho$ that allows for the spreading of the contagion process. Thus there are two parameters underlying the mobility dynamics that we can either hold fixed or let free. In Fig. 3 we show the phase diagram in the $\sigma$-$\tau$ space separating the global invasion from the extinction regime. The phase diagram tells us that,  all parameters being equal, the rate of diffusion to nearby subpopulations has to be larger than  $\sigma_c$ to guarantee the spreading of the contagion. Analogously, if we allow $\tau$ to vary, we observe that the global spreading of the contagion process can be achieved by extending the visit times $\tau$ of individuals in nearby subpopulations above a definite threshold $\tau_c$. The explicit expressions of the threshold values can be found in the Supplementary Information.  

Another very interesting feature of the above threshold condition is the explicit effect of the network topology encoded in the moments of the degree distribution. Indeed, the heterogeneity of the network favors the global spread of the contagion process by lowering the threshold value. In the Supplementary Information we show that in the case of heavy-tailed degree distribution the threshold virtually reduces to zero for infinitely large system sizes. Even at finite size, however, the threshold value is generally smaller for networks with greater heterogeneity as is shown in Fig. 3, which compares the phase diagrams of heterogeneous and homogeneous networks of same size. 
In order to test the validity of the analytical picture obtained here, we have performed an extensive set of Monte Carlo numerical simulations of the contagion process in large subpopulation networks. The simulations are individual based and consider the commuting and contagion dynamics microscopically with no approximations as detailed in the Supplementary Information. The substrate network is given by an uncorrelated random complex network~\cite{Molloy1998} generated with the uncorrelated configuration model~\cite{Catanzaro2005} to avoid inherent structural correlations.  In Fig. 3 we report the results for a network with Poissonian degree distribution and a network with power-law degree distribution $P(k)\sim k^{-2.1}$. Individuals are distributed heterogeneously in each subpopulation according to the relation $N_k={\overline N} k / \langle k \rangle$, where ${\overline N}=10^4$. Although the analytical phase diagram has been derived by using several approximations, it matches the numerical simulations qualitatively and quantitatively, as shown by the good agreement of the analytical phase boundary and the numerical simulations in Fig.~3b. We also report in Fig.~ 4a the behavior of the number of invaded populations as a function of commuting rates. The phase transition between the invasion and extinction regimes at a specific value of $\rho=\sigma \tau$ is clearly observed in the microscopic simulations. 

\noindent{\bf Data-driven simulations}.

\noindent As a further confirmation of the validity of the theoretical results we have tested our results in a real-world setting. We have considered the commuting network of all counties in the continental US as obtained by the US Census 2000 data~\cite{USCensusBureau}. In this dataset each subpopulation represents a county and a connection the presence of commuting flow between two counties. In the simulation each county is associated with its actual population and each link with a specific commuting rate from the real data. We have considered only short-range commuting flows up to 125 miles. The visit time has been considered to be of the order of a working day (8 hours). On this real data layer we have simulated the spreading of an SIR contagion process and studied the number of infected counties as a function of the global rescaling factor of the commuting rates. 
It is remarkable to observe that in the case of the real data a clear phase transition exists between the two regimes  at a critical value of the global rescaling factor of the commuting rates. In Fig.~ 4 we also illustrate the different behavior of the contagion process in the two regimes by mapping the number of infected counties in the US as a function of time.

\noindent{\bf Conclusions}.

\noindent While the presented results are anchored upon the example of disease spread, the metapopulation approach can be abstracted to the phenomena of knowledge diffusion, online community formation, information spread, and technology. In all these examples, we have individuals stationed primarily in well-defined subpopulations, with occasional interactions with other subpopulations governed by interaction rates similar in scheme to those presented here. While most of the studies in defining epidemic threshold have focused on single populations, it is clear that more attention must be devoted to the study of the spread in structured populations. In this case the understanding of the invasion threshold is crucial to the analysis of large-scale spreading across communities and subpopulations. The theoretical approach presented in this paper opens the path to the inclusion of more complicated mobility or interaction scheme and at the same time provides a general framework that may be used not just as an interpretative framework but a quantitative and predictive framework as well. Understanding the effect of mobility and interaction patterns on the global spread of contagion processes can indeed be used to devise enhanced or suppressed spread by acting on the basic parameters of the system in the appropriate way, which might find applications ranging from the protection against emerging infectious diseases to viral marketing.

\newpage
\noindent
{\large {\bf Methods}}

\small {
\noindent
{\bf Stationary populations.} Rate equations characterizing the commuting dynamics among subpopulations can be defined by using the variables
$N_{kk}(t)$ and $N_{kk'}(t)$ as
\begin{eqnarray}
\partial_t N_{kk}(t) &=& -\sigma_k N_{kk}(t) + \tau^{-1} k \sum_{k'} N_{kk'}(t) P(k'|k) \;\;\;, \\
\partial_t N_{kk'}(t) &=& \sigma_{kk'} N_{kk}(t) - \tau^{-1} N_{kk'}(t) \;\;\;,
\end{eqnarray}
where $\sigma_{kk'}$ is the rate at which an individual of subpopulation $k$ commutes to neighboring subpopulation $k'$. Then, considering 
the statistical equivalence of subpopulations with the same degree and the mean field assumption we have $\sigma_k=k \sum_{k'} \sigma_{kk'} P(k'|k)$
where $P(k'|k)$ is the conditional probability of having a subpopulation $k'$ in the neighborhood of a subpopulation $k$. Equilibrium is given by the condition $\partial_tN_{kk}=\partial_tN_{kk'}=0$ and yields the relation 
\begin{equation}
N_{kk'}=N_{kk} {\sigma_{kk'}  \tau} \;\;\;.\label{equilibrium}
\end{equation}
Using the expression $N_k=N_{kk}(t)+k \sum_{k'} N_{kk'}(t) P(k'|k)$ for total number of individuals of subpopulation $k$ one can obtain the stationary populations in equations (\ref{eq:N_kk}) and (\ref{eq:N_kk'}).

\smallskip
\noindent
{\bf Branching process.} Each  subpopulation of degree $k'$ invaded by the contagion process at the $n-1$th generation may seed its $k'-1$ neighbors at most (all of its neighbors minus the one from which it got the infection). 
The probability of finding a subpopulation of degree $k$ in the neighborhood is $P(k|k')$. For each neighboring subpopulation, 
the probability that it has not already been invaded by the contagion process in an earlier generation is $\prod_{m=0}^{n-1} (1-D_k^m/V_k)$. 
If $\lambda_{k'k}$ infectious seeds are sent to the neighbor,  the outbreak occurs with probability $1-R_0^{-\lambda_{k'k}}$~\cite{Bailey1975}.
We can then relate the number of diseased subpopulations at the $n$th generation with that at the $n-1$th generation as the simultaneous realization of all these above conditions, 
\begin{equation}
\begin{array}{rl}
D_{k}^{n}=  \sum_{k'}D_{k'}^{n-1}(k'-1)\left[1-R_{0}^{-\lambda_{k'k}}\right]P(k|k') \prod_{m=0}^{n-1}\left(1-\frac{D_{k}^{m}}{V_{k}}\right)\;\;\;.
\end{array}
\end{equation}
In the early stage of the contagion process we can assume that  $\prod_{m=0}^{n-1} (1-D_k^m/V_k)\simeq 1$. We will also consider the case that we are just above the local epidemic threshold, $R_0-1 \ll 1$, 
so that the outbreak probability can be approximated by $1-R_{0}^{-\lambda_{k'k}} \simeq (R_0-1)\lambda_{k'k}$. If we also ignore degree correlations between neighboring subpopulations, $P(k|k')=kP(k) / \langle k \rangle$~\cite{Barrat2008},
we obtain equation (\ref{branch}).

\smallskip
\noindent
{\bf Invasion threshold.} In order to obtain the explicit  expression for the subpopulation reproductive number in equation (\ref{eq:InvasionThr1}) we need to derive an expression for $\lambda_{k'k}=\left(N_{k'k}+N_{kk'}\right)\alpha$. This expression depends on  the form of commuting rates among  subpopulations. We consider the case in which
\begin{equation}
\sigma_{kk'} = \sigma {N_{k'} \over N_k + N_k^{\rm nn} } \;\;\;,
\end{equation}
where $N_k^{\rm nn}=k \sum_{k'} N_{k'} P(k'|k)$ is the average total population in the neighborhood of subpopulation $k$.
The above expression assumes that the per-capita mobility rate is rescaled by the number of individuals in the subpopulation~\cite{Keeling2002}, thus leading to $\sigma_k$ that decreases as $N_k$ increases. 
This behavior account for the effect introduced by large subpopulation sizes; the overall per capita commuting rate outside of the subpopulation generally decreases in large populations as individuals tend to commute internally. 
In this case we obtain
\begin{equation}
\sigma_{kk'} = \sigma {\langle k \rangle k' \over (\langle k \rangle+ \langle k^2 \rangle) k} \;\;\;.
\end{equation}
This expression allows the calculation of $N_{kk'}$ and using  the approximate relation for the fraction of infected cases generated by the end of the SIR epidemic~\cite{Keeling2008} introduced into a fully susceptible population $\alpha \simeq 2 (R_0-1) / R_0^2$, we obtain the expression for $\lambda_{k'k}$:
\begin{equation}
\lambda_{k'k} = {2 \overline{N} (R_0-1) \rho \over R_0^2 \langle k^2 \rangle (1+ \langle k \rangle / \langle k^2\rangle+\rho)} (k'+k)\;\;\;.
\end{equation}
If we substitute the above relation into equation (\ref{branch}) we get
\begin{eqnarray}
D_{k}^{n}=\frac{2 \overline{N} (R_{0}-1)^2 \rho }{R_{0}^{2} \langle k^{2}\rangle\langle k\rangle (1+\langle k \rangle / \langle k^2\rangle+\rho) } kP(k) \sum_{k'}D_{k'}^{n-1}(k'-1)\left(k+k'\right)\;\;\;. 
\end{eqnarray}
In order to write a closed form of the above iterative process we introduce the definitions $\Theta_0^n\equiv \sum_k  (k-1) D_k^n$ and 
$\Theta_1^n\equiv \sum_k  k(k-1) D_k^n$ whose next generation equations are defined as
\begin{equation}
{\bf \Theta^{n}}=G{\bf \Theta^{n-1}\;\;\;{\rm with}}\;\;\;{\bf \Theta^{n}}=\left(\begin{array}{c}
\Theta_{0}^{n}\\
\Theta_{1}^{n}\end{array}\right) \;\;\;,
\end{equation}
where $G$ is a $2\times2$ matrix,
\begin{equation}
G=   \frac{2 \overline{N} (R_{0}-1)^2 \rho }{R_{0}^{2} \langle k^{2}\rangle\langle k\rangle (1+\langle k \rangle / \langle k^2\rangle+\rho) }
\left(\begin{array}{cc}
\langle k^{3}\rangle-\langle k^{2}\rangle &   \langle k^{2}\rangle-\langle k\rangle \\ 
\langle k^{4}\rangle-\langle k^{3}\rangle &   \langle k^{3}\rangle-\langle k^{2}\rangle\end{array}\right) 
\label{eq:GenerationOp} \;\;\;.
\end{equation} 
The global behavior of the contagion process across the network of subpopulations is determined by the largest eigenvalue $R_\ast$ of $G$ as expressed in equation ($\ref{eq:InvasionThr1}$) where $F$ is a function of 
the moments of degree distribution,
\begin{eqnarray}
F (\langle k\rangle, \langle k^{2}\rangle, \langle k^{3}\rangle,\langle k^{4}\rangle) 
\equiv
{1\over \langle k\rangle \langle k^{2}\rangle} \Bigr[ \langle k^{3}\rangle-\langle k^{2}\rangle  + \left(\langle k^{4}\rangle-\langle k^{3}\rangle\right)^{1/2}\left(\langle k^{2}\rangle-\langle k\rangle\right)^{1/2} \Bigr] \;\;\;. \label{eq:ftopology}
\end{eqnarray}
}


\bigskip
\noindent
{\large {\bf Acknowledgments}}\\
We would like to thank to Chiara Poletto and Vittoria Colizza for interesting discussions during the preparation of this manuscript. 
This work has been partially funded by the NIH R21-DA024259 award and the DTRA-1-0910039 award to AV; The work has been also partly sponsored by the Army Research Laboratory and was accomplished under Cooperative Agreement Number W911NF-09-2-0053. The views and conclusions contained in this document are those of the authors and should not be interpreted as representing the official policies, either expressed or implied, of the Army Research Laboratory or the U.S. Government. 

\noindent
{\large {\bf Author Contributions}}\\
D.B. and A.V. have conceived and executed the study, performed the analytical calculations and drafted the manuscript. D.B. has performed the numerical simulations. 

\newpage
\begin{figure}[!ht]
\begin{center}
\includegraphics[width=16cm]{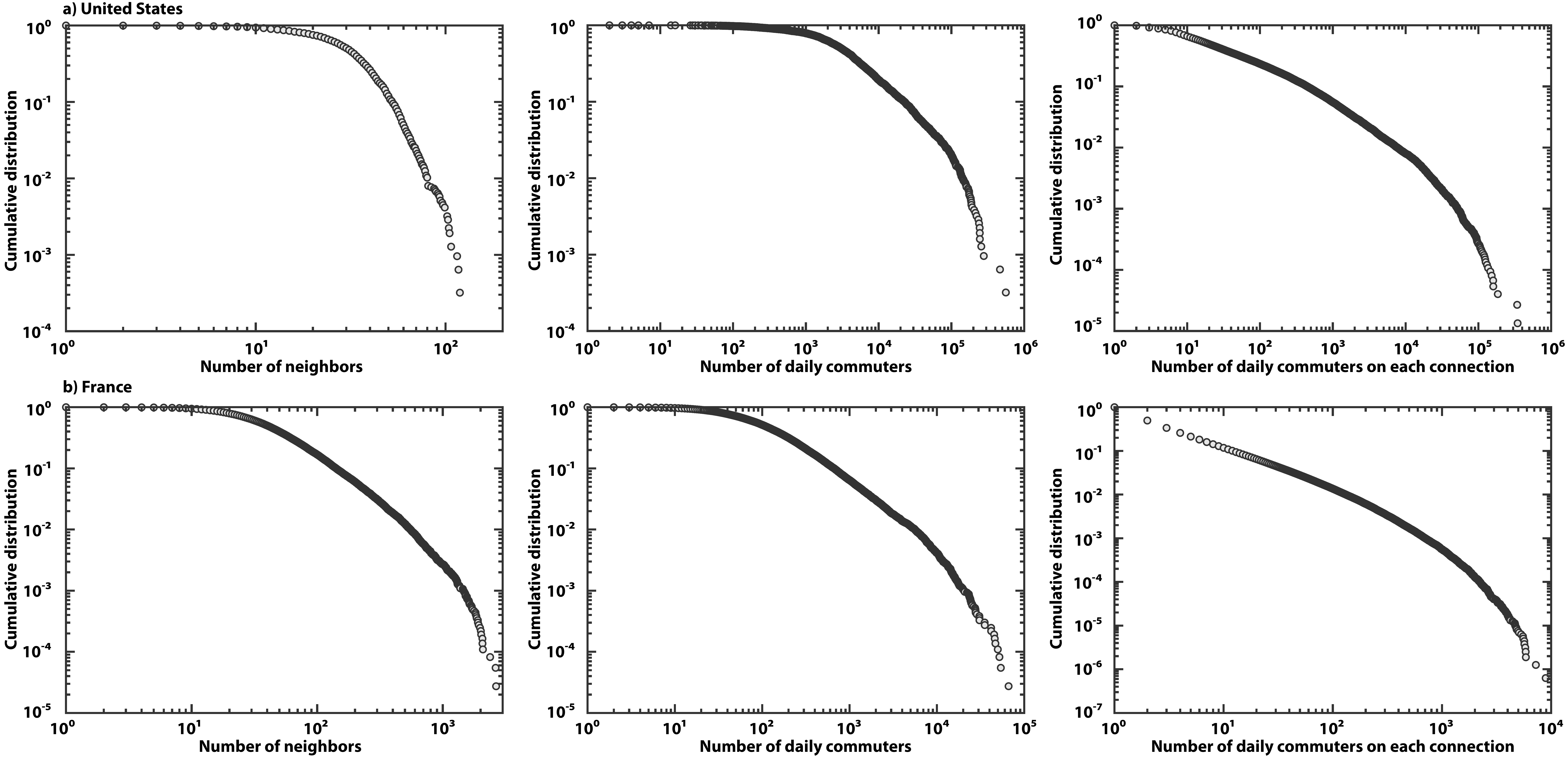}
\vskip .1cm 
\caption{\small {\bf Statistical properties of commuting networks in the United States and France.} {\bf a,} Commuting network in the United States at the level of counties (http://www.census.gov/). {\bf b,} Commuting network in France at the level of municipalities (http://www.insee.fr/). Cumulative distributions of the number of connections (left) and the number of daily commuters (center) per administrative unit, as well as the number of daily commuters on each connection (right) are displayed. The networks are highly heterogeneous in the number of connections as well as in the commuting fluxes.}
\label{fig:Fig1}
\end{center}
\end{figure}

\newpage
\begin{figure}[!ht]
\begin{center}
\includegraphics[width=16cm]{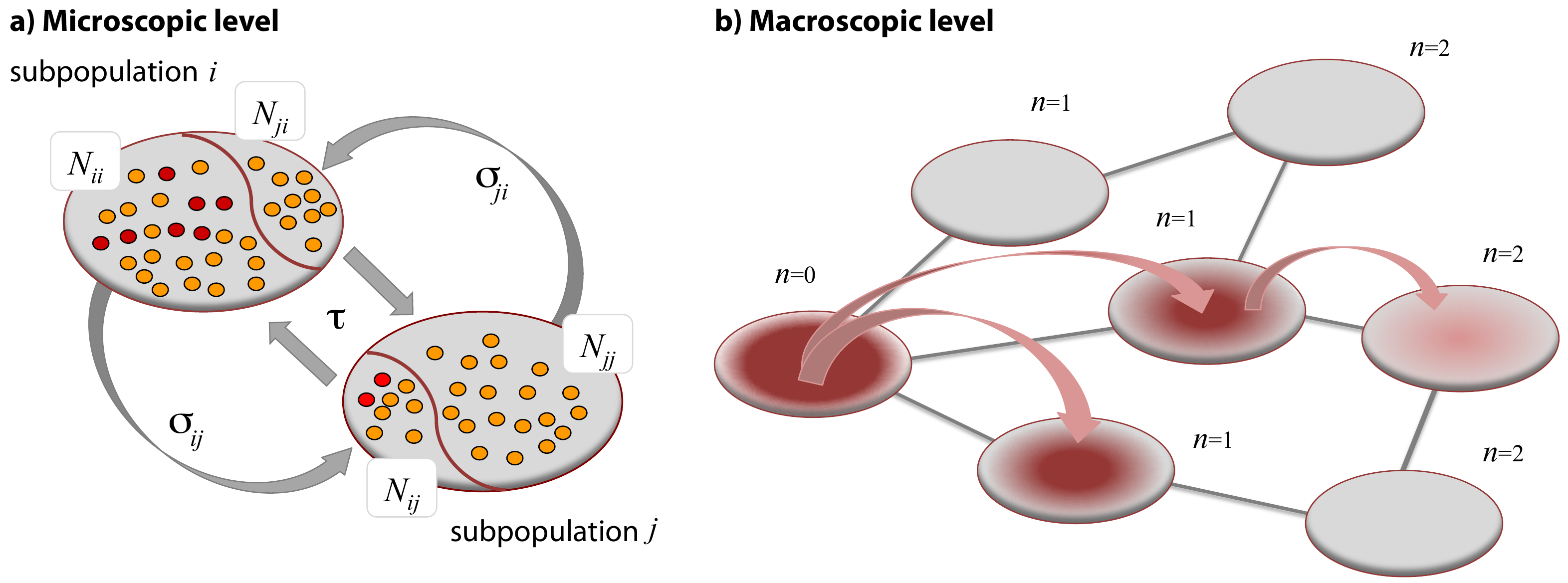}
\vskip .1cm 
\caption{\small {\bf Illustration of the subpopulation invasion dynamics.}
{\bf a,} Mixing of two subpopulations and contagion dynamics due to commuting at the microscopic level. 
At any time subpopulation $i$ is occupied by a fraction of its own population $N_{ii}$ and a fraction of individuals $N_{ji}$ whose origin is in neighboring subpopulation $j$.
The figure depicts the flux of individuals back and forth between the two subpopulations due to  commuting process. This exchange of individuals is the origin of the transmission of the contagion process  from subpopulation $i$ to subpopulation $j$. The contagion process  is mediated by contacts between infectious (red particles) and susceptible (yellow particles) individuals. {\bf b,} Macroscopic representation of invasion dynamics. Nodes are organized from left to right according to their generation index $n$. Arrows indicates the transmission of the contagion process from a diseased subpopulation at the $n-1$th generation to a subpopulation at the $n$th generation.}
\label{fig:Fig2}
\end{center}
\end{figure}

\newpage
\begin{figure}[!ht]
\begin{center}
\includegraphics[width=16cm]{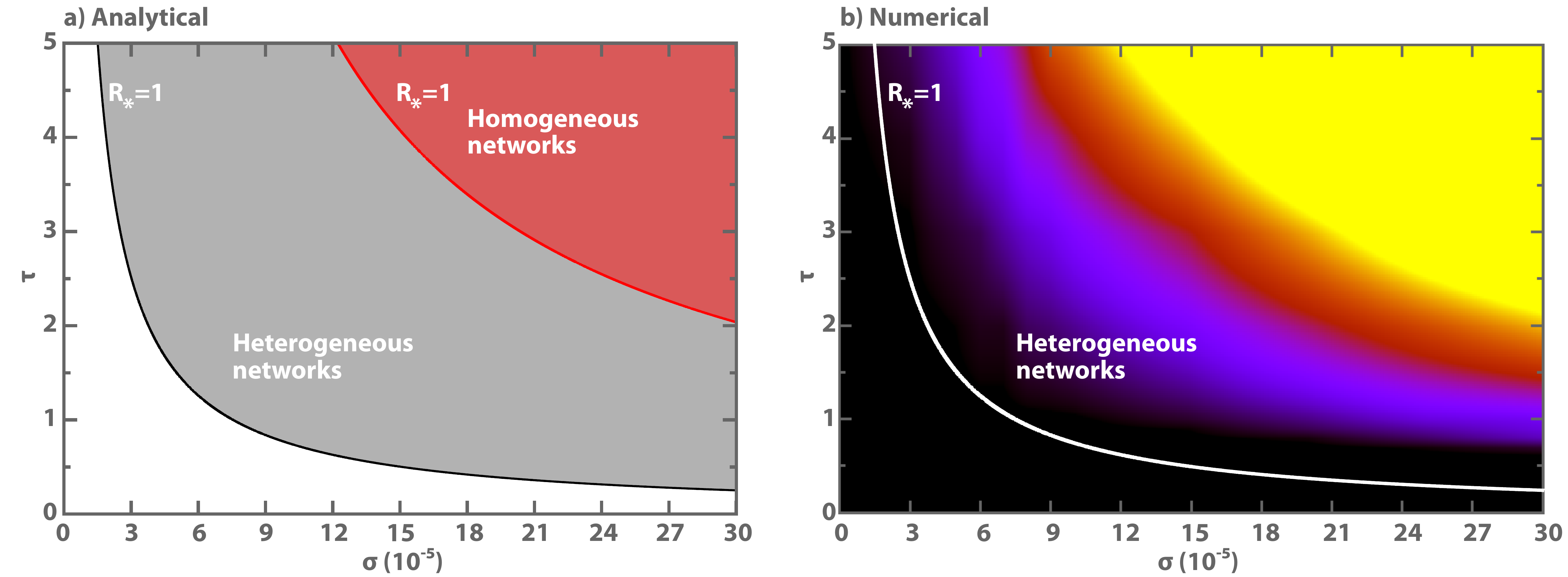}
\vskip .1cm 
\caption{\small {\bf Phase diagrams separating the global invasion regime from the extinction regime.}
{\bf a,} Plot of the equation (\ref{eq:InvasionThr1}) in the $\sigma$-$\tau$ space.
The red and black lines identify the $R_\ast=1$  relation for the homogeneous and  heterogeneous uncorrelated random networks, respectively.
The global spreading regime is in the region of parameters indicated by shaded areas. 
The networks are made of $V=10^4$ subpopulations, each of which accommodates  a degree dependent population of $N_k={\overline N} k / \langle k \rangle$ individuals, with ${\overline N}=10^4$. Both networks have the same average degree 
in which the heterogeneous network has degree distribution $P(k)\sim k^{-2.1}$ and the homogeneous network has Poisonian degree distribution. The SIR dynamics is characterized by $R_0=1.25$ and $\mu^{-1}=15 \;days$.
{\bf b,} Numerical simulations on heterogeneous networks. The system assumes the same parameter values of ({\bf a}). Color scale from black to yellow is
linearly  proportional to the number of infected subpopulations. Black indicates an invasion of less than 0.1\% of subpopulations and yellow indicates an invasion of more than 10\% of subpopulations.}
\label{fig:Fig3}
\end{center}
\end{figure}

\newpage
\begin{figure}[!ht]
\begin{center}
\includegraphics[width=16cm]{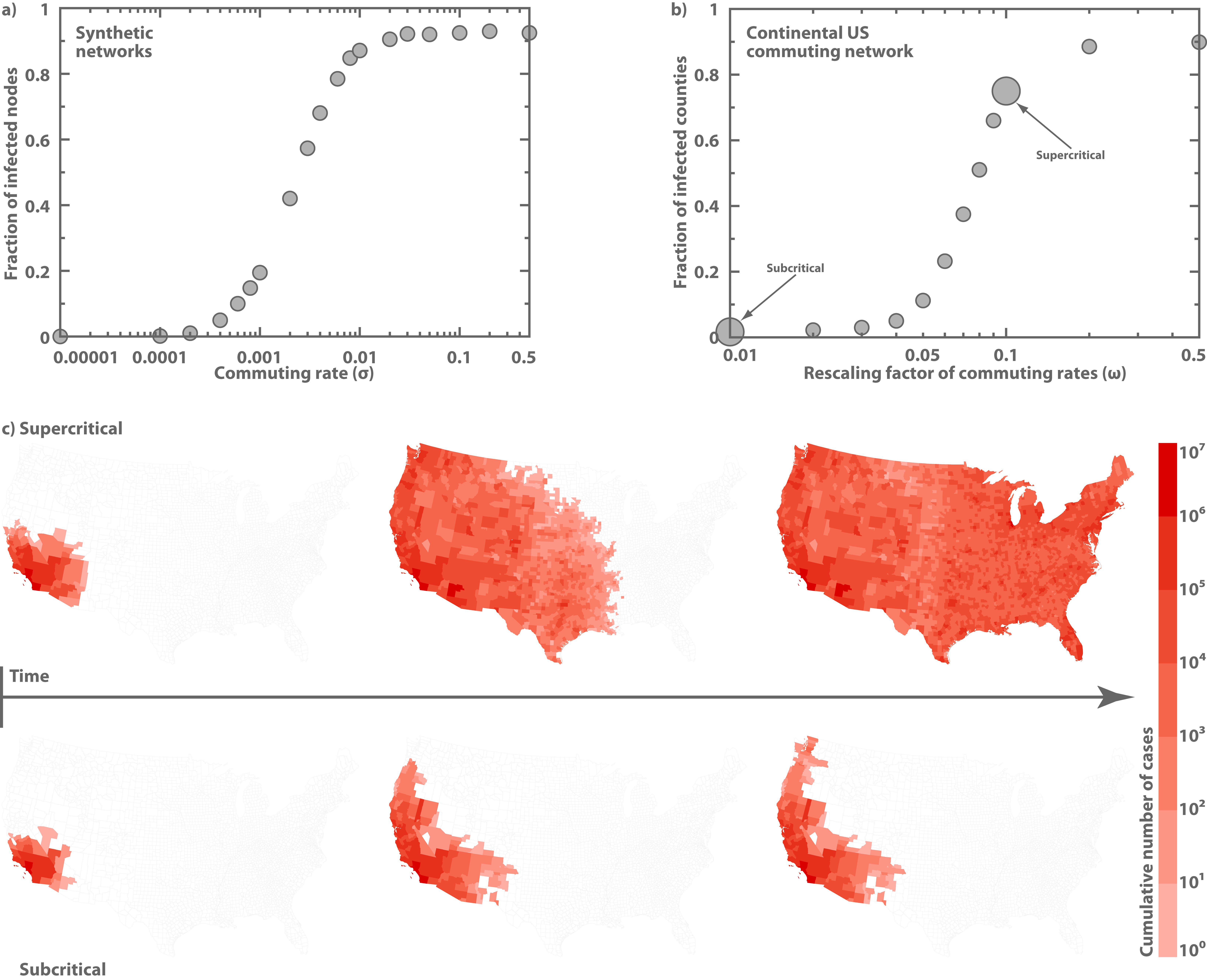}
\vskip .1cm 
\caption{\small {\bf Dynamical behavior of an SIR epidemic on the real US commuting network data.}
{\bf a,} Average fraction of infected subpopulation as a function of commuting rates in networks with the same statistical properties as the heterogeneous network in Fig.~3a.
Visit time in this case is fixed at $\tau=1 \; day$.  
{\bf b,} Average fraction of infected subpopulations as a function of the intensity of commuting fluxes in the US. We study the system behavior by varying all commuting rates  $\sigma_{ij}$ between county pairs by a factor $\omega$ as $\sigma_{ij}\to\omega \sigma_{ij}$. Visit time assumes a realistic value of $\tau= 8 \; hours$. The infection is initially seeded in Los Angeles. The data considers only real commuting flows up to 125 miles and the actual county populations (see text). 
{\bf c,} Temporal progression of average cumulative number of infected cases in the subcritical and supercritical regimes of the invasion dynamics. 
The rescaling factors used in these simulations are marked in ({\bf b}). The SIR dynamics assumes $R_0=1.25$ and $\mu^{-1}=3.6\;days$ in both cases.}
\label{fig:Fig4}
\end{center}
\end{figure}

\clearpage
\newpage
\noindent
{\large \large  \bf Supplementary Information}\\\\ 
\noindent
{{\large \bf Invasion Threshold}}\\
The global behavior of the contagion process is determined by the largest eigenvalue $R_\ast$ of the subpopulation next generation matrix $G$ as detailed in the Methods section of main paper.
If the eigenvalue $R_\ast>1$ we have that the subpopulation invasion process is supercritical and the disease will be able to globally spread across subpopulations. This is equivalent to define a subpopulation reproductive number $R_\ast$~\cite{Ball1997, Cross2005, Cross2007, Colizza2007PRL,Colizza2008JTB} that in structured metapopulation systems is equivalent to basic reproductive number $R_0$ at the single population level: 
\begin{equation}
R_{\ast} = \frac{2 \overline{N} (R_{0}-1)^2 \rho }{R_{0}^{2} (1+  \langle k \rangle / \langle k^2 \rangle + \rho )} F(\langle k\rangle, \langle k^{2}\rangle, \langle k^{3}\rangle,\langle k^{4}\rangle)  
\;\;\;, 
\end{equation}
where
\begin{eqnarray}
&F (\langle k\rangle, \langle k^{2}\rangle, \langle k^{3}\rangle,\langle k^{4}\rangle) \equiv
{1\over \langle k\rangle \langle k^{2}\rangle} \Bigr[ \langle k^{3}\rangle-\langle k^{2}\rangle  + \left(\langle k^{4}\rangle-\langle k^{3}\rangle\right)^{1/2}\left(\langle k^{2}\rangle-\langle k\rangle\right)^{1/2} \Bigr] \;\;\;. 
\end{eqnarray}
The infectious diseases will spread globally in the metapopulation system only if $R_{\ast}>1$. 
Thus, by setting $R_\ast=1$, we can define an epidemic threshold relation for the mobility ratio $\rho$,
\begin{eqnarray}
\rho_c = {1 + \langle k \rangle / \langle k^2 \rangle \over 2 \overline{N} (1-R_{0}^{-1})^2 F (\langle k\rangle, \langle k^{2}\rangle, \langle k^{3}\rangle,\langle k^{4}\rangle)  -1} \;\;\;,
\end{eqnarray}
below which the infection remains confined to a small number of subpopulations. In an infinite metapopulation system the threshold is defined rigorously and the fraction of infected subpopulations is zero below the threshold and finite only if the mobility parameters set the ratio $\rho$ above the threshold value. 
The threshold value is defined for the ratio between the rates characterizing the mobility process. This condition is therefore twofold on the mobility dynamics if we fix one of the two parameters $\sigma$ and $\tau$, and let the other parameter free. On one hand the threshold relation is $\sigma_c = \rho_c \; \tau^{-1}$,
\begin{equation}
\sigma_c = { (1 + \langle k \rangle / \langle k^2 \rangle)  \tau^{-1} \over 2 \overline{N} (1-R_{0}^{-1})^2 F (\langle k\rangle, \langle k^{2}\rangle, \langle k^{3}\rangle,\langle k^{4}\rangle)  -1} \;\;\;.
\end{equation}
This intuitively states that the rates of diffusion has to be large enough ($\sigma > \sigma_c$) to guarantee the spreading of the disease. 
Interestingly, however, we can also define the threshold relation for  $\tau$ by $\tau_c= \rho_c \sigma^{-1}$,
\begin{equation}
\tau_c = {  (1 + \langle k \rangle / \langle k^2 \rangle) \sigma^{-1} \over 2 \overline{N} (1-R_{0}^{-1})^2 F (\langle k\rangle, \langle k^{2}\rangle, \langle k^{3}\rangle,\langle k^{4}\rangle)  -1} \;\;\;, \end{equation}
that is telling that the global spreading of the contagion porcess can be achieved by reducing the return rates of individuals; in other words by extending the visit times of individuals in nearby subpopulations ($\tau > \tau_c$). This last conditions however breaks down when $\tau$ becomes much larger than the contagion time scale thus breaking the time-scale separation~\cite{Keeling2002} assumption used here.

Another very interesting feature of the above threshold condition is the explicit effect of the network topology encoded in the moments of degree distribution $\langle k \rangle$, $\langle k^2 \rangle $, etc. As already been observed in the case of Markovian diffusion case~\cite{Colizza2007PRL,Colizza2008JTB}, the heterogeneity of the network favors the global spread of the epidemic by lowering the threshold value. 
Indeed, for heavy-tailed degree distribution $P(k)\sim k^{-\gamma}$ with $\gamma > 1$, the $n$th moment scales as $k_{\rm max}^{n+1-\gamma}$ if $ n \ge  \gamma-1$ and $k_{\rm max}\gg k_{\rm min}$.
This means that for $n \ge \gamma-1$, the $n$th  moment of the degree distribution tends to diverge in the infinite size limit of the network, as in this limit $k_{\rm max} \rightarrow \infty$, virtually reducing the threshold to zero. Even at finite size, however, the threshold value is generally smaller the higher the network heterogeneity is. 
In order to make this last statement transparent, we will turn our attention to the scaling of the moments of degree distribution for very large system sizes. In the case of  $1 < \gamma < 2$, the term $\langle k \rangle / \langle k^2 \rangle$ in the nominator of $\rho_c$
scales as
\begin{equation}
{\langle k \rangle \over \langle k^2 \rangle} \sim k_{\rm max}^{-1} \;\;\;.
\end{equation}
In the range $2 < \gamma < 3$, the scaling is
\begin{equation}
{\langle k \rangle \over \langle k^2 \rangle} \sim k_{\rm max}^{\gamma-3} \;\;\;.
\end{equation}
In all the other cases of $\gamma > 3$, the term $\langle k \rangle / \langle k^2 \rangle$ has a finite value. That means that the nominator of $\rho_c$ is finite for any $\gamma > 1$.
Now lets turn our attention to the denominator of $\rho_c$ and analyze the scaling of $F(\langle k\rangle, \langle k^{2}\rangle, \langle k^{3}\rangle,\langle k^{4}\rangle)$. 
In the case of $1 < \gamma < 2$, $F$ scales as 
\begin{eqnarray}
F(\langle k\rangle, \langle k^{2}\rangle, \langle k^{3}\rangle,\langle k^{4}\rangle)\sim { \langle k^{3}\rangle + \langle k^{4}\rangle^{1/2}\langle k^{2}\rangle^{1/2}  \over \langle k\rangle \langle k^{2}\rangle} 
\sim k_{\rm max}^{\gamma-1} \;\;\;.
\end{eqnarray}
In the case of $2 < \gamma < 3$, second moment $\langle k^{2} \rangle$ in the denominator and higher moments in the numerator are dominant, leading to the scaling relation:
\begin{eqnarray}
F(\langle k\rangle, \langle k^{2}\rangle, \langle k^{3}\rangle,\langle k^{4}\rangle)\sim { \langle k^{3}\rangle + \langle k^{4}\rangle^{1/2}\langle k^{2}\rangle^{1/2}  \over \langle k^{2}\rangle} 
\sim k_{\rm max} \;\;\;.
\end{eqnarray}
In the range $3 < \gamma < 4$, only the third moment $\langle k^{3}\rangle$ in the numerator dominates, thus
\begin{eqnarray}
F(\langle k\rangle, \langle k^{2}\rangle, \langle k^{3}\rangle,\langle k^{4}\rangle)\sim  \langle k^{3}\rangle 
\sim k_{\rm max}^{4-\gamma} \;\;\;.
\end{eqnarray}
In the range $4 < \gamma < 5$, only the fourth moment $\langle k^{4}\rangle$ in the numerator dominates, leading to
\begin{eqnarray}
F(\langle k\rangle, \langle k^{2}\rangle, \langle k^{3}\rangle,\langle k^{4}\rangle)\sim  \langle k^{4}\rangle^{1/2} 
\sim k_{\rm max}^{(5-\gamma)/2} \;\;\;.
\end{eqnarray}
The above expressions state that for any heavy-tailed degree distribution with exponent $\gamma < 5$, $F (\langle k\rangle, \langle k^{2}\rangle, \langle k^{3}\rangle, \langle k^{4}\rangle)$ tents to diverge in the limit of infinite network size, which in turn pushes the threshold value $\rho_c$ to zero. While, on the other hand, if $\gamma > 5$ then  $F(\langle k\rangle, \langle k^{2}\rangle, \langle k^{3}\rangle, \langle k^{4}\rangle)$ has a finite value.
\\ \\ 
\noindent
{\large {\bf Computational model}}\\
Inside each subpopulation we consider an SIR epidemic model~\cite{Keeling2008}, in which each individual is classified by one of the discrete disease states at any point in time. 
The rate at which a susceptible person in subpopulation $i$ acquires the infection, the so-called force of infection~\cite{Keeling2008} $\lambda_i$,  is determined by interactions with infectious individuals.
The force of infection $\lambda_i$ acting on each susceptible individual in subpopulation $i$ has been assumed to follow the mass action principle
\begin{equation}
\lambda_i (t) = \beta {I_i^\ast(t) \over N_i^\ast(t)}\;\;\;,
\end{equation}
where $\beta$ is the transmission rate of infection and $I_i^\ast(t)/N_i^\ast(t)$ is the prevalence of infectious individuals in the subpopulation. 
Each person in the susceptible compartment (S) contracts the infection with probability $\lambda_i(t) \Delta t$ and enters the infectious compartment (I), where $\Delta t$ is the time interval considered. Each infectious individual permanently recovers with probability $\mu \Delta t$, entering the recovered compartment (R).

\noindent
{\bf  Synthetic  subpopulation networks}. \\
\smallskip
\noindent
{\bf \it Generation of substrate networks}. In order to compare with theoretical calculations, topologically uncorrelated random graphs have been considered.
In this case, analytical calculations show that epidemic invasion threshold only depends on the degree distribution of the subpopulation networks.
In order to verify this result, two different network topologies have been generated:
\begin{itemize} 
\item
Erd\H{o}s-R\'{e}nyi graphs~\cite{Erdos1959} have been synthetized by assigning a link between each pair of nodes with probability $\langle k\rangle/(V-1)$, where $V$ is the number of nodes and $\langle k\rangle$ is a prescribed average node degree.
\item
Networks with power-law degree distribution, $P(k)\sim k^{-\gamma}$ with $k_{\rm min} \le k\le k_{\rm max}$, have been generated by uncorrelated configuration model~\cite{Molloy1998,Catanzaro2005}.
All the scale-free networks have been generated by setting $\gamma=2.1$ and $k_{\rm min}=2$. 
\end{itemize}
For the sake of comparison, the average degree of Erd\H{o}s-R\'{e}nyi graphs has been set to that of scale-free networks. 

\smallskip
\noindent
{\bf \it Subpopulation sizes}. From a pool of $\overline{N} V$ people, a population size $N_i$ is assigned to each subpopulation $i$, defining its permanent residents.
The population size is chosen at random from a multinomial distribution with probability proportional to $k_i$, which
ensures the metapopulation system to obey $N_k = {\overline{N} k / \langle k \rangle}$.

\smallskip
\noindent
{\bf \it Mobility parameters}. The rate $\sigma_{ij}$ at which a resident of subpopulation $i$ commutes to a neighboring subpopulation $j\in\upsilon(i)$ assumes 
\begin{equation}
\sigma_{ij} = \sigma {N_j \over N_i + \sum_{\ell \in \upsilon(i)} N_\ell} \;\;\;.
\end{equation}
Each resident in subpopulation $i$ leaves its origin and visits subpopulation $j$ with probability $\sigma_{ij} \Delta t$.
A commuter in subpopulation $j$ returns back to its resident subpopulation $i$ with probability $\tau^{-1} \Delta t$.

Simulations have been initialized with $I(0)=10$ infectious individuals, seeded randomly in a single subpopulation of degree $k_{{\rm min}}$, while the rest of the population is assumed to be susceptible to infection. 

\noindent
{\bf Real-world subpopulation networks}. \\
Realistic simulations have been performed using the county to country commuting network in the continental United States~\cite{USCensusBureau}. 
The network consists of about $3,100$ nodes, each of which corresponds to a US county. Weighted link from node $i$ to node $j$ represents 
the daily number of commuters from county $i$ to county $j$. Thus, the population size of each node and the commuting rates among them are fully determined by the
data. The return rate $\tau^{-1}$, however, has been set to $\tau^{-1}=3\;day^{-1}$ corresponding to a regular working day (8 hours). 
Simulations have been initialized with $I(0)=10$ infectious individuals seeded in Los Angeles County, California.

\noindent
{\bf Statistical analysis}.\\
Since we aim at determining the epidemic invasion threshold, we have let the metapopulation system run until the infection dies out.
In the numerical results presented in main paper, all the realizations resulting in at least one diseased subpopulation have contributed to the statistical analysis. For each set of parameters, we have generated at least $1,000$ system realizations. Since the subpopulation networks
and dynamical processes on them are subject to fluctuations in the case of synthetic populations, we have sampled at least $10-20$ network realizations and $100-200$ dynamical realizations on each of them. While in the case of the real-world scenarios, we have generated at least $1,000$ dynamical realizations.

\noindent
{\bf Contagion and mobility dynamics}. \\
We will follow the notations defined in main paper and represent each individual by its disease state $X$, its permanent subpopulation $i$ and its present subpopulation $j\in\upsilon(i)$. 
Since all the individuals sharing the same three indices $\left(X,\;i,\;j\right)$ are identical in terms of the dynamical processes, we are going refer to the number of such individuals at time $t$ by $X_{ij}(t)$. 
Then, by definition, the instantaneous compartment size $X_j^\ast(t)$ in subpopulation $j$ can be expressed as
\begin{eqnarray}
X_j^\ast (t)&=& X_{jj}(t) + \sum_{\ell \in \upsilon(j)} X_{\ell j} (t) \;\;\;,
\end{eqnarray}
and the total number of individuals as $N_j^\ast = \sum_X X_{j}^\ast$.
The number of individuals in each compartment $X$ with a residence in $i$ and present in $j$ is subject to discrete and stochastic  dynamical processes
defined by disease and transport operators. The disease operator  ${\cal D}_{j}$ represents the change due to the compartment transition induced by the infection dynamics, and the transport operator $\Omega_{X}$ represents the variation due to mobility.
\\
The term ${\cal D}_j$ can be written as a combination of a set of transitions $\{{\cal D}_j(X,Y)\}$, where ${\cal D}_j(X,Y)$ represents the number of transitions from compartment $X$ to $Y$ and is simulated as an integer random number extracted 
from a multinomial distribution. Then the change due to infection dynamics reads as
\begin{equation}
{\cal D}_j(X)=\sum_{Y} \left[ {\cal D}_j(Y,X)-{\cal D}_j(X,Y) \right] \;\;\;.
\end{equation}
As a concrete example let us consider the temporal change in the infectious compartment. There is only one possible transition from the compartment, that is to the recovered compartment.
The number of transitions is extracted from the binomial distribution
\begin{equation}
{\rm Pr}^{\rm Binom}(I_{ij}(t), p_{ I_{ij} \rightarrow R_{ij} })\;\;\;,
\end{equation}
determined by the transition probability
\begin{eqnarray}
p_{ I_{ij}\rightarrow R_{ij} } & = & \mu \Delta t\;\;\;,
\end{eqnarray}
and the number of individuals in the compartment $I_{ij}(t)$ (its
size). This transition causes a reduction in the size of the
compartment. The increase in the compartment size is due to
the transitions from susceptible into infectious compartment. This is also a random
number extracted from the binomial distribution 
\begin{equation}
{\rm Pr}^{\rm Binom}(S_{ij}(t),p_{{S_{ij}\rightarrow I_{ij}}}) \;\;\;, 
\end{equation}
given by the chance of contagion 
\begin{eqnarray}
p_{{S_{ij}\rightarrow I_{ij}}} & = & \lambda_{j}(t) \Delta t\;\;\;,
\end{eqnarray}
and the number of attempts equal to the number of susceptibles $S_{ij}(t)$.
After extracting these numbers from the appropriate distributions,
we can calculate the total change ${\cal D}_j(I)$ in infectious compartment as 
\begin{eqnarray}
{\cal D}_j (I)={\cal D}_{j}(S,I)-{\cal D}_{j}(I,R)\;\;\;.
\end{eqnarray}
\\
Transport operator $\Omega_X$ expresses the total change in compartment sizes due to the commuting of permanent residents of subpopulation $i$ back and forth. 
The variation in $X_{ij}$ can be decomposed into $\Omega^\rightarrow_X(i,j)$ and $\Omega^\leftarrow_X(j,i)$ as
\begin{equation}
\Omega_X= \Omega^\rightarrow_X(i,j) - \Omega^\leftarrow_X(j,i) \;\;\;.
\end{equation}
The first term $\Omega^\rightarrow_X(i,j)$ represents an increase that is caused by the departing residents of subpopulation $i$ to visit subpopulation $j$.
The $\Omega^\rightarrow_X(i,j)$ is a random number extracted from the multinomial distribution
\begin{equation}
{\rm Pr}^{\rm Multinom}(X_{ii}(t),\{p_{{X_{ii}\rightarrow X_{i\ell} }} | \ell \in \upsilon(i)\}) \;\;\;, 
\end{equation}
determined by the probability of commuting to subpopulation $j$
\begin{equation}
p_{{X_{ii}\rightarrow X_{ij} }} = \sigma_{ij} \Delta t \;\;\;, 
\end{equation}
and the number of such trails $X_{ii}(t)$.
The second term $\Omega^\leftarrow_X(j,i)$ corresponds to a reduction in $X_{ij}$ and is due to the return trips from subpopulation $j$ to permanent subpopulation $i$. 
The $\Omega^\leftarrow_X(j,i)$ is also a random number extracted from the binomial distribution 
\begin{equation}
{\rm Pr}^{\rm Binom}(X_{ij}(t),p_{X_{ij}\rightarrow X_{ii}}) \;\;\;, 
\end{equation}
given by the probability of returning home
\begin{equation}
p_{X_{ij}\rightarrow X_{ii}} = \tau^{-1} \Delta t \;\;\;, 
\end{equation}
and the size of the compartment $X_{ij}(t)$.
\\
We have assumed that the infection does not alter people's behavior, i.e., all the compartments are identical in their mobility.


\begin{thebibliography}{50}

\bibitem{Marro1999} Marro, J. \& Dickman, R. {\it Nonequilibrium Phase Transitions in Lattice Models} (Cambridge Univ. Press, Cambridge, 1999).

\bibitem{vanKampen1981} van Kampen, N. G. {\it Stochastic Processes in Physics and Chemistry} (North-Holland, Amsterdam, 1981).

\bibitem{May1984} May, R. M. \& Anderson, R. M. Spatial heterogeneity and the design of immunization programs. {\it Math. Biosci.} {\bf 72,} 83$-$111 (1984).

\bibitem{Bolker1993} Bolker, B. M. \& Grenfell, T.  Chaos and biological complexity in measles dynamics. {\it Proc. R. Soc. London B} {\bf 251,} 75$-$81 (1993).

\bibitem{Bolker1995}Bolker, B. M. \& Grenfell, T.  Space persistence and dynamics of measles epidemics. {\it Philos. Trans. R. Soc. London B} {\bf  348,} 309$-$320 (1995).

\bibitem{Sattenspiel1995} Sattenspiel, L. \& Dietz, K.  A structured epidemic model incorporating geographic mobility among regions. {\it Math. Biosci.} {\bf 128,} 71$-$91 (1995).

\bibitem{Lloyd1996}Lloyd, A. L. \& May, R. M. Spatial heterogeneity in epidemic models. {\it J. Theor. Biol.} {\bf 179,} 1$-$11 (1996).

\bibitem{Keeling2002} Keeling, M. J. \& Rohani, P. Estimating spatial coupling in epidemiological systems: a mechanistic approach. {\it Ecol. Lett.} {\bf 5,} 20$-$29 (2002).

\bibitem{Watts2005} Watts, D., Muhamad, R., Medina, D. C. \& Dodds, P. S. Multiscale
resurgent epidemics in a hierarchical metapopulation model. {\it Proc. Natl. Acad. Sci. USA} {\bf 102,} 11157$-$11162 (2005).

\bibitem{Rapoport1953} Rapoport, A. Spread of information through a population with socio-structural bias: I. assumption of transitivity. 
{\it Bull. Math. Biol.} {\bf 15,} 523$-$533 (1953).

\bibitem{Goffman1964} Goffman, W. \& Newill, V. A. Generalization of epidemic eheory: An application to the transmission of ideas. 
{\it Nature} {\bf 204,} 225 $-$ 228 (1964).

\bibitem{Goffman1966} Goffman, W. Mathematical approach to the spread of scientific ideas$-$the history of mast cell research. {\it Nature} {\bf 212,} 449$-$452 (1966).

\bibitem{Dietz1967} Dietz, K. Epidemics and rumours: A survey. {\it J. of Royal Stat. Soc. A} {\bf 130,} 505$-$528 (1967).

\bibitem{Tabah1999}  Tabah, A. N. Literature dynamics: Studies on growth, diffusion, and epidemics. 
{\it Ann. Rev. Inform. Sci. Technol.} {\bf 34,} 249$-$286 (1999).

\bibitem{Daley2000} Daley, D. J. \& Gani, J. {\it Epidemic Modeling: An Introduction} (Cambridge Univ. Press, Cambridge, 2000).

\bibitem{Rvachev1985} Rvachev, L. A. \& Longini, I. M.   A mathematical model for the global spread of influenza. {\it Math. Biosci.} {\bf 75,} 3$-$22 (1985).

\bibitem{Grais2003} Grais R. F., Hugh Ellis, J. \& Glass, G. E.   Assessing the impact of airline travel on the geographic spread of pandemic influenza. 
{\it Eur. J. Epidemiol.} {\bf 18,} 1065$-$1072 (2003).

\bibitem{Hufnagel2004} Hufnagel L., Brockmann D. \& Geisel T.  Forecast and control of epidemics in a globalized world. 
{\it Proc. Natl. Acad. Sci. USA} {\bf  101,} 15124$-$15129 (2004).

\bibitem{Colizza2006a} Colizza, V., Barrat, A., Barth\'{e}lemy, M. \& Vespignani, A.  The role of the airline transportation network in the prediction and predictability of global epidemics. {\it Proc. Natl. Acad. Sci. USA} {\bf 103,} 2015$-$2020 (2006).

\bibitem{Balcan2010} Balcan, D. {\it et al.}
Modeling the spatial spread of infectious diseases: The GLobal Epidemic and Mobility computational model. {\it Journal of Computational Science} {\bf 1,} 132$-$145 (2010). 

\bibitem{Chowell2003} Chowell, G., Hyman, J. M., Eubank, S. \& Castillo-Chavez, C. Scaling laws for the movement of people between locations in a large city. 
{\it Phys. Rev. E} {\bf 68,} 066102 (2003).

\bibitem{Barrat2004} Barrat, A., Barth\'{e}lemy, M., Pastor-Satorras, R. \& Vespignani, A. The architecture of complex weighted networks. {\it Proc. Natl. Acad. Sci. USA} {\bf 101,} 3747$-$3752 (2004).

\bibitem{Guimera2005} Guimer\'{a}, R.,	Mossa,	S.,	Turtschi,	A. \&	Amaral,	L. A. N.		
The worldwide air transportation network: anomalous centrality, community structure, and cities' global roles. {\it Proc. Natl.  Acad. Sci. USA} {\bf 102,} 7794$-$7799 (2005).

\bibitem{Brockmann2006} Brockmann D., Hufnagel L. \&  Geisel T.  The scaling laws of human travel. {\it Nature} {\bf 439,} 462$-$465 (2006).

\bibitem{Patuelli2007} Patuelli R., Reggiani R., Gorman S. P., Nijkamp P. \& Bade F. -J. Network analysis of commuting flows: A comparative static approach to German data. {\it Networks Spatial Econ.} {\bf 7,} 315$-$331 (2007).

\bibitem{Gonzalez2008} Gonz\'{a}lez M. C., Hidalgo C. A. \& Barab\'{a}si A. -L.  Understanding individual human mobility patterns. {\it Nature} {\bf 453,} 779$-$782 (2008).

\bibitem{Wang2009} Wang P. \&  Gonz\'{a}lez, M. C.   Understanding spatial connectivity of individuals with non-uniform population density.
{\it Phil. Trans. R. Soc. A} {\bf 367,} 3321$-$3329 (2009).

\bibitem{Song2010a} Song, C., Qu, Z., Blumm,  N. \& Barab\'{a}si, A. -L.    Limits of Predictability in Human Mobility. {\it Science} {\bf 327,} 1018$-$1021 (2010).

\bibitem{Song2010b} Song, C., Koren, T., Wang, P. \& Barab\'{a}si, A. -L. Modelling the scaling properties of human mobility. {\it Nat. Phys.} {\bf 6,} 818$-$823 (2010).

\bibitem{Balcan2009-PNAS} Balcan, D. {\it et al.}  Multiscale mobility networks and the spatial spreading of infectious diseases.
{\it Proc. Natl.  Acad. Sci. USA} {\bf 106,} 21484$-$21489 (2009).

\bibitem{Colizza2007NatPhys} Colizza, V., Pastor-Satorras, R. \& Vespignani, A. Reaction-diffusion processes and metapopulation models in heterogeneous networks. 
{\it Nat. Phys.} {\bf 3,} 276$-$282 (2007).

\bibitem{Colizza2007PRL} Colizza, V. \& Vespignani, A.  Invasion threshold in heterogeneous metapopulation networks. {\it Phys. Rev. Lett.} {\bf 99,} 148701 (2007).

\bibitem{Colizza2008JTB} Colizza, V. \& Vespignani, A. Epidemic modeling in metapopulation systems with heterogeneous coupling pattern: Theory and simulations. {\it J. Theor. Biol.} {\bf 251,} 450$-$467 (2008).

\bibitem{barthelemy2010} Barth\'{e}lemy, M., Godr\`{e}che, C. \& Luck, J.-M. Fluctuation effects in metapopulation models: percolation and pandemic threshold. {\it J. Theor. Biol.} {\bf 267,} 554$-$64 (2010).

\bibitem{Ni2009} Ni, S. \& Weng, W. Impact of travel patterns on epidemic dynamics in heterogeneous spatial metapopulation networks. {\it Phys. Rev. E} {\bf 79,} 016111 (2009).

\bibitem{Ben-Zion2010} Ben-Zion, Y., Cohena, Y. \& Shnerba, N.M. Modeling epidemics dynamics on heterogenous networks.
{\it J. Theor. Biol.} {\bf 264,} 197$-$204 (2010).

\bibitem{Pastor-Satorras2001} Pastor-Satorras, R. \& Vespignani, A. Epidemic spreading in scale-free networks. {\it Phys. Rev. Lett.} {\bf 86,} 3200$-$3203 (2001).

\bibitem{lloyd2001} Lloyd, A. L. \& May, R. M. How viruses spread among computers and people. {\it Science} {\bf 292,} 1316$-$1317 (2001).

\bibitem{Cohen2003} Cohen, R., Havlin, S. \& ben-Avraham, D. Efficient immunization strategies for computer networks and populations.  {\it Phys. Rev. Lett.} {\bf 91,} 247901 (2003).

\bibitem{Barrat2008}  Barrat, A., Barth\'{e}lemy, M. \& Vespignani, A.  {\it Dynamical Processes on Complex Networks} (Cambridge Univ. Press, Cambridge, 2008).

\bibitem{Keeling2008} Keeling, M. J. \& Rohani, P.  {\it Modeling Infectious Diseases in Humans and Animals}  (Princeton Univ. Press, Princeton, 2008).  

\bibitem{Ball1997} Ball, F., Mollison, D. \& Scalia-Tomba, G.  Epidemics with two levels of mixing. {\it Ann. Appl. Probab.} {\bf 7,} 46$-$89 (1997).

\bibitem{Cross2005} Cross, P., Lloyd-Smith, J. O., Johnson, P. L. F. \& Wayne, M. G. Duelling timescales of host movement and disease recovery determine invasion of disease in structured populations. {\it Ecol. Lett.} {\bf 8,} 587$-$595 (2005).

\bibitem{Cross2007} Cross, P., Johnson, P. L. F., Lloyd-Smith, J. O. \& Wayne, M. G. Utility of $R_0$ as a predictor of disease invasion in structured populations. {\it J. R. Soc. Interface} {\bf 4,} 315$-$324 (2007).

\bibitem{Harris1989} Harris, T. E. {\it The Theory of Branching Processes} (Dover, New York, 1989).

\bibitem{Vazquez2006} V\'{a}zquez, A.  Polynomial growth in age-dependent branching processes with diverging reproductive number. 
{\it Phys. Rev. Lett.} {\bf 96,} 038702 (2006).

\bibitem{Molloy1998} Molloy, M. \& Reed, B.  The size of the largest component of a random graph on a fixed degree sequence. 
{\it Combinatorics, Probab. Comput.} {\bf 7,} 295$-$306 (1998).

\bibitem{Catanzaro2005} Catanzaro, M., Bogun\~{a}, M. \& Pastor-Satorras, R.  Generation of uncorrelated random scale-free networks. {\it Phys. Rev. E} {\bf 71,} 027103 (2005).

\bibitem{USCensusBureau} U.S. Census Bureau, 
http://www.census.gov/.

\bibitem{Bailey1975} Bailey, N. T. {\it The Mathematical Theory of Infectious Diseases} (Macmillan, New York, 1975).

\bibitem{Erdos1959} Erd\H{o}s, P. \& R\'{e}nyi, A.  On random graphs. {\it Publ. Math.} {\bf 6,} 290$-$297 (1959).

\end{thebibliography}
\end{document}